\documentclass[fleqn,usenatbib]{mnras}

\usepackage{newtxtext,newtxmath}

\usepackage[T1]{fontenc}
\usepackage{ae,aecompl}

\usepackage{graphicx}	
\usepackage{amsmath}	
\usepackage{amssymb}	
\usepackage{xcolor}

\title[SNR in Multiphase ISM]{Evolving Supernova Remnants in Multiphase Interstellar Media}

\author[M. A. Villagran et al.]{
Marco A. Villagran,$^{1}$\thanks{E-mail: mvillagran@iafe.uba.ar}
P.F. Vel\'azquez,$^{2}$
D.O. G\'omez,$^{1,3}$
E.B. Giacani$^{1,4}$
\\
$^{1}$CONICET-Universidad de Buenos Aires, Instituto de Astronom\'{i}a y F\'{i}sica del Espacio, Ciudad Universitaria, Bs. As., Argentina\\
$^{2}$Instituto de Ciencias Nucleares, UNAM, Circuito Exterior s/n, Ciudad Universitaria, Ciudad de M\'exico, M\'exico\\
$^{3}$Departamento de F\'isica, Facultad de Ciencias Exactas y Naturales, Universidad de Buenos Aires, Bs. As., Argentina\\
$^{4}$Facultad de Arquitectura, Dise\~no y Urbanismo, Universidad de Buenos Aires, Bs. As., Argentina\\
}

\date{Accepted XXX. Received YYY; in original form ZZZ}

\pubyear{2019}

\hypersetup{draft}
\begin{document}
\label{firstpage}
\pagerange{\pageref{firstpage}--\pageref{lastpage}}
\maketitle

\begin{abstract}
We performed three-dimensional magnetohydrodynamic simulations to study the evolution of a supernova remnant (SNR) in a turbulent neutral atomic interstellar medium. The media used as background shares characteristics with the Solar neighbourhood and the SNR has mass and energy similar to those of a Type Ia object. Our initial conditions consist of dense clouds in a diluted medium, with the main difference between simulations being the average magnitude of the magnetic field. We measured amplifications of the magnetic energy of up to 34$\%$ and we generated synthetic maps that illustrate how the same object can show different apparent geometries and physical properties when observed through different lines of sight.        
\end{abstract}




\section{Introduction}
A key element in the vast zoology of astronomical objects are supernova (SN) explosions, they are events that deposit close to $10^{51}$\,erg of mechanic energy into the interstellar medium (ISM) and are intrinsically related to the multiscale dynamics of galaxies. These phenomena also enrich the interstellar medium  as they increase its metallicity. They also inject thermal energy through ejecta that moves with speeds close to the $10^4$\,km\,s{$^{-1}$}, a crucial ingredient for the turbulent nature of the gas. After an explosion of this kind a supernova remnant (SNR) is left. SNRs come in a huge variety of sizes and shapes, they also radiate in different wavelengths. The differences that exist between them may be due to several factors, like the environment in which they explode and the mechanism that induced the explosion. 
 
According to the explosion mechanism, SNe can be classified into two groups: Types Ib, Ic, and II are the product of gravitational core-collapse of massive stars (M $\geq 8 M_{\odot}$ in the main sequence) that have exhausted all their nuclear fuel. Depending on the mass of the stellar core, it collapses leaving either a neutron star or a black hole.
The remnants produced by gravitational collapse of massive stars, due to their short lifetimes, are found very close or immersed in their parental molecular cloud.
The other group, supernova Type Ia,  occurs in a binary system, where an old degenerate star, thought to be an exploding carbon-oxygen white dwarfs with mass close to the Chandrasekhar limit (1.38 M$_{\odot}$), accretes mass to the point of surpassing said limit. The two main models for the acquisition of said extra material is through the merger of two white dwarfs (called the double degenerate scenario) or from the the accretion of the material of a companion star (the single degenerate scenario). In this case, the explosion destroys the star completely \citep[See for SNR reviews the works of][]{Reynolds12,Vink2016}. In general, Type Ia SNe live in a depleted low density media, which has been characterized with densities between $1$\,cm$^{-3}$ and $5$\,cm$^{-3}$ \citep[see][]{Yamaguchi14}.

The Solar neighbourhood normally refers to the region that is near the Sun. It is a low density region ($n_{sn} \sim 2\,$cm$^{-3}$ compared to molecular clouds with typical densities $n_{mc} \sim 100\,$cm$^{-3}$) populated mainly by red dwarf stars, main sequence M-type stars, and to a lower extent, white dwarfs. These stars are embedded in an ISM that is mainly atomic and segregated in two thermally stable phases (with an important fraction of the gas as thermally unstable). This segregation results from the thermal instability (TI) proposed by \cite{Field65}, and it gives rise to cold clump-like structures in an attenuated warm gas. These two phases have similar pressure and are caused by the balance of the energy acquired via star radiation and cosmic rays and the one lost by line-emission \citep{Wolfire95}. The TI is said to be driven by turbulence \citep[see ][]{Field65} and it requires an input of thermal and kinetic energy. The main candidates for this energy supply are the supernova explosions via the interaction of SNRs with the ISM \citep{McKee77,Cox81}.

The interaction between a SNR and the ISM is not easy to study analytically, as the inhomogeneous nature of the medium poses a challenging problem. The lion's share of the modeling efforts have been done for a SNR expanding on a uniform medium, see \citet{Sedov59} or \citet{Reynolds2016} and the references therein. Efforts have also being made to tackle the inhomogeneous medium case, like the self similar solutions for a SNR expanding in a medium with evaporating clouds described in \citet{White91} or the papers tackling the mass-loading scenario for SNR in clumpy media \citep[e.g.][]{Dyson02}. In addition, it is also difficult to obtain observational data of these interactions and it has the additional problem of not knowing the state of the ISM previous to the explosion. Nevertheless the relevance of the properties of the ISM are widely accepted, as the pre-SN density distribution can alter the observed shape of the SNR \citep[e.g.][]{Giacani09} and even the magnetic field distribution may affect how a SNR is observed \citep[e.g.][]{Gaensler98}. Auxiliary tools for studying this phenomena come in the form of numerical simulations. An important amount of modelling and testing has been done evolving SNRs in a uniform medium, see for example: \citet{Chevalier74} and \citet{Cioffi88}. More recently, some groups have published their results of simulations that evolve SNRs on in-homogeneous media. For example: \citet{Martizzi15} or \citet{Zhang2019}, who focused their study on media more akin to dense molecular clouds. In the diluted media side of the spectrum, the work done by \citet{Inoue09} presents two-dimensional simulations of the evolution of SNR in a two-phase gas. 

In this paper we test the expansion of SNRs in a two-phase atomic auto-consistent ISM via three dimensional magnetohydrodinamical (MHD) simulations and study dynamical and synthetically observed results. In section \ref{sec:Model} we describe our model end the conditions that we tested. Next in section \ref{sec:Results} we present our results. In section \ref{sec:Discussion} we present a discussion of our data. Finally in section \ref{sec:Conclusions} we list our conclusions.

\section{Numerical Model and Parameters}
\label{sec:Model}

For the present work we simulated the approximate evolution of a SNR that was created via a Type Ia SN scenario using a set of eulerian MHD simulations. For our purposes we worked in three main steps:  
\begin{itemize}
  \item Simulate the atomic ISM;
  \item Evolve the SNR in the generated media; 
  \item Create synthetic observation maps from the results.
\end{itemize}
In this section we outline the aforementioned steps.

\subsection{Initial conditions of the SNR and the ISM}

 For the first step we used initial conditions similar to the density, velocity, temperature and magnetic field of the Solar neighbourhood. The initial ISM are the ones described by \citet{Villagran18} in which the authors evolved the atomic ISM up to the moment where turbulence is well developed and the gas is adequately segregated by the means of \textit{Thermal Instability} (TI) \citep[see][]{Field65} parting from a different set of initial conditions that are originally in an unstable thermal equilibrium. Those simulations were performed using an Eulerian scheme on a fixed mesh in a periodic cube of $L = 100\,$pc using $512^3$ points with a continuous source of energy in the form of a single wave number (corresponding to $l \sim 50\,$pc) solenoidal external forces on Fourier space. The original purpose of those simulations was to test the effects of different initially uniform magnetic fields in the ISM formed by the TI. The details of the numerical scheme used to prepare the ISM are summarized in \citet{Gazol16}. 

The typical density distribution of the atomic ISM is a bimodal distribution with peaks in the dense ($50$ cm$^{-3}$) and in the diluted ($0.5$ cm$^{-3}$) regions, those peaks correspond to two different temperatures, a cold temperature ($T \sim 100\,$K) for the dense clumps and a warm one ($T \sim 8000\,$K)  for the diluted background. The magnetic field intensities and root mean square velocity ($v_{\mathrm{rms}}$) distributions are heavily centred at a single value with dispersion that are typically small. For the velocity distribution the central value is at $v_{\mathrm{rms}} \sim 7\,$km$\,$s$^{-1}$ while the magnetic field distributions are centred at $B \sim 1.2\,$$\mu$G or $B \sim 9\,$$\mu$G. A velocity distribution as the one described before is typical for a turbulent neutral atomic ISM. For observational details of the atomic ISM see for example \citet{Kulkarni87}, \citet{Dickey90}, \citet{Dickey02}, and \citet{Kalberla09}. The shock front of the SNR affects the values of those physical quantities modifying the distributions in time. As described by the so called \textit{Free Expansion} and the \textit{Sedov} phases it is expected that the mass in the vicinity of the shock front keeps increasing via sweeping the ISM, hence modifying the density distribution of the region. Similar changes are expected for the temperature, magnetic field intensity (\cite{Giacalone07}) and v$_{\mathrm{rms}}$ distributions.   

Here after we abbreviate these simulations as BLT for the lowest magnetic field intensity ($B \sim 1.2\,$$\mu$G) and as BHT for the highest magnetic field intensity ($B \sim 9\,$$\mu$G), see table\,\ref{tab:initial_values}. 

\subsection{Evolution of the SNR}

From the set of beforehand described data cubes we selected cubic $L = 25\,$pc regions with the same total mass ($M_T \sim 620\,$M$_\odot$) which we fed as initial conditions to the GUACHO MHD code that allocated and evolved a SNR with typical conditions of initial energy $E_0 \sim 10^{51}\,$erg in the form of ejected mass ($M_0 \sim 30\,$M$_\odot$) with velocities $v_0 \sim 10^4\,$km$\,$s$^{-1}$. The energy is distributed so that
95 $\%$ of it is kinetic and the remaining 5 $\%$ is thermal. The smaller cube was then extrapolated from $128^3$ points to $512^3$ points which results in a resolution of $\sim 0.05$\,pc. This allowed us to resolve in better detail the SNR at small time scales ($t \sim 100\,$yr) while maintaining the inhomogeneous signature of the ISM. For an in-depth description of GUACHO we refer to \citet{villarreal18}. 

In addition to the aforementioned simulations we also ran a pair of fiducial tests, one for each mean magnetic field used. The main characteristics of these tests is that they start from a uniform, isothermal and static field, whose densities and temperatures are equal to the mean initial values of the main simulations and with a uniform magnetic field along a single axis with an intensity equal to the mean initial magnetic fields of the turbulent set.  This pair of runs will be referred as BLU ($B \sim 1.5\,$$\mu$G) and BHU ($B \sim 9\,$$\mu$G), see table\,\ref{tab:initial_values}. 

\begin{figure}
	\includegraphics[width=\columnwidth]{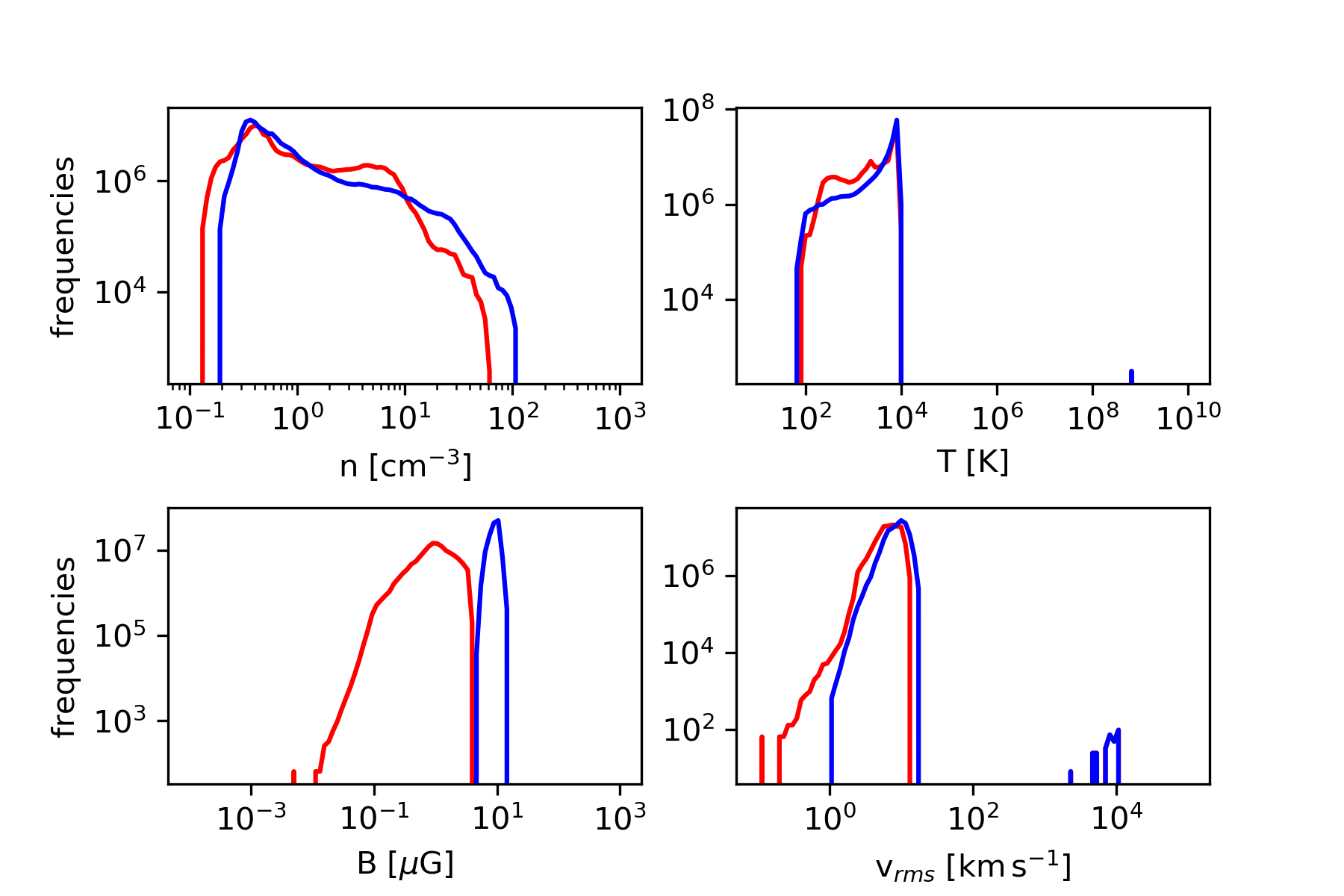}
    \caption{Histograms for the initial distributions of physical quantities of interest. From top left in clockwise direction: density, temperature, v$_{\mathrm{rms}}$ and magnetic field intensity. The colors represent both turbulent simulations, BLT (red) and BHT (blue).}
    \label{fig:initial_conditions}
\end{figure}

\begin{table}
\caption{Table containing the initial maximum value of density, temperature, magnetic field and velocity for all the simulations.}
\centering
	\begin{tabular}{|c c c c c|} 
 \hline
  Simulation & n [cm$^{-3}$] & T [K] & B$_{\mathrm{rms}}$ [$\mu$G] & v$_{\mathrm{rms}}$ [km s$^{-1}$] \\ 
 \hline
 BLT & 59 & 10035 & 4 & 14.3 \\ 
 \hline
 BHT & 107 & 9954 & 15 & 18.8 \\
 \hline
 BLU & 1.7 & 1585 & 1.5 & 0 \\
 \hline
 BHU & 1.7 & 2202 & 9 & 0 \\
\end{tabular}
\label{tab:initial_values}
\end{table}

The density, temperature, magnetic field intensity and $v_{\mathrm{rms}}$ initial distributions are shown in fig. \ref{fig:initial_conditions}. The maximum values of these distributions can be found in table \ref{tab:initial_values}. These values were used to measure how much the interaction with the shock front alters the general distribution's behaviour and as a base for the measured amplifications of the physical quantities.  

\subsection{Synthetic Observations of the SNR}

For the sake of comparison with observational results, we estimated the emitted radiation originating in the different components of our MHD simulations, projected it into a plane and applied an absorption factor due to an assumed column density. The material associated to the SNR has a temperature high enough for its thermal emission peak to be in the $[0.5, 10]$~keV X-ray range. The interaction of a young SNR with the ISM can also trigger the formation of dust particles that can radiate infrared light and polarize the radio emission coming from the gas. This radio emission near the blast waves corresponds to the synchrotron effect suffered by the electrons interacting with the locally amplified magnetic field. The synchrotron emission, in extraordinary cases, has a spectrum that extends all the way up to the X-ray regime. However, such a process requires instabilities able to generate particle acceleration and magnetic field amplification, which clearly go beyond the MHD description and clearly lie outside the scope of this work since our resolution is larger than the greater Larmor radius for a thermal proton in every simulation.      
We focused our study on the radio and thermal X-ray emission since they are some of the most relevant spectral regions in young SNR observations. 

With the purpose of describing the polarized radiation, from our numerical results we calculate the total intensity of the synchrotron emission and the Stokes parameters $Q$ and $U$ by using \citep{Schneiter15,Cecere16}:

\begin{equation}
I(x,y)=\int_{LoS} i(\nu) dz
\label{I}
\end{equation}

\begin{equation}
Q(x,y)\propto \int_{LoS} F_p i(\nu){\cos{2\chi}} dz
\label{Q}
\end{equation}

\begin{equation}
U(x,y)\propto \int_{LoS} F_p i(\nu)\sin{2\chi} dz
\label{U}
\end{equation}
being $x$ and $y$ the coordinates in the plane of the sky, $z$ is the coordinate along the line of sight (LoS),  $\chi$ is the local position angle of the magnetic field (measured counterclockwise with respect to $y-$axis, on the plane of the sky), $F_p$ is the polarization factor (equal to $(\alpha+1)/(\alpha+5/3)$), and $i(\nu)$ is the synchrotron emissivity at a frequency $\nu$, which is given by \citep{Cecere16}:

\begin{equation}
    i(\nu)= C_2 \rho v^{4\alpha} B_\perp^{\alpha+1} \nu^{-\alpha},
    \label{inu}
\end{equation}
being $\rho$ the gas density, $v$ its velocity, $\alpha$ the spectral index\footnote{It is assumed that the relativistic electrons have a power law spectrum given by $N(E)=K\ E^{-p}$, where $p$ is the electron spectral index, which  is related with   $\alpha$ as $\alpha=(p-1)/2$.}, $B_\perp$ is the component of the magnetic field in the plane of the sky, and $C_2$ if a factor depending on the obliquity angle $\theta_B$ (the angle between the magnetic field and the normal shock) calculated as follow \citep{Reynolds98}:

\begin{equation}
C_2= K_2 (cos^2\theta_B+ \frac{1}{1+\eta^2} sin^2\theta_B)
\label{c2}
\end{equation}
where $K_2$ is a constant and $\eta$ is a parameter which indicates if the magnetic field is ordered ($\eta >> 1$) or not ($\eta << 1$). Following \'Avila-Aroche et al. (in preparation) this parameter is estimated as:
\begin{equation}
    \eta = \bigg(\frac{<B_{\perp}>}{\sigma_{B_{\perp}}}\bigg)^2,
    \label{etastat}
\end{equation}
being $<B_{\perp}>$ and $\sigma_{B_{\perp}}$ the mean value and the standard deviation of the magnetic field in the plane of the sky, respectively. It is important to highlight that quasi-parallel ($\propto \cos^2{\theta_B}$) and quasi-perpendicular ($\propto \sin^2{\theta_B}$) acceleration mechanisms are present in Eq.(\ref{etastat}).

Similarly from our numerical simulations, synthetic thermal X-ray emission maps were calculated,
considering the energy range [0.1-10] keV, by integrating the X-ray
emissivity $j=n^2 \phi(T)$ (low-density regime) along the line of sight
($n$ and $T$ are the density and the temperature, respectively. The
function $\phi(T)$ was calculated by using the CHIANTI database
\citep{dere97,landi2006}, considering solar abundances and the
ionization equilibrium of \citet{mazzotta98}.

\section{Results}
\label{sec:Results}

\subsection{Results from Direct Measurements}

\begin{figure}
	\includegraphics[width=\columnwidth]{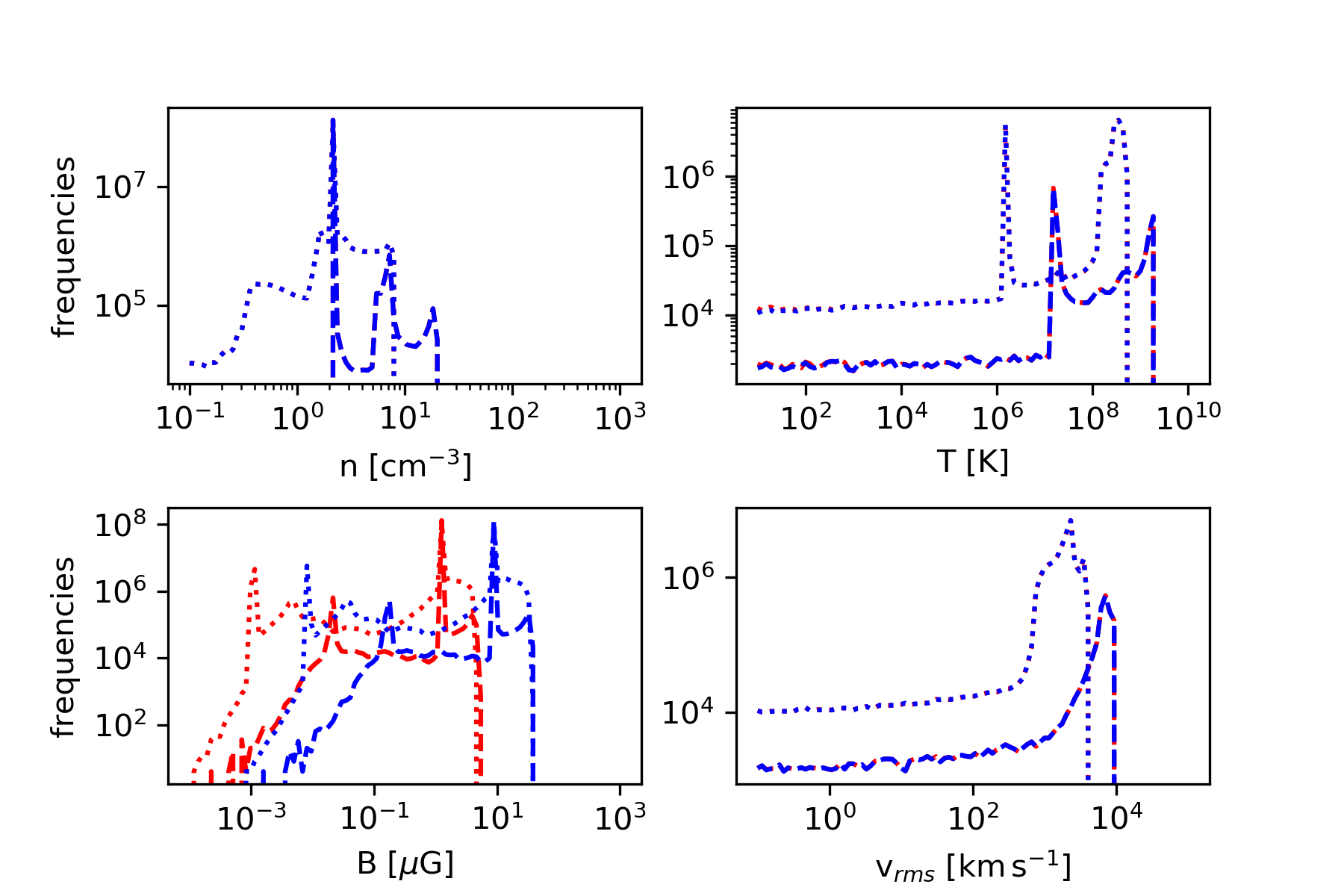}
    \caption{Histograms for the initial distributions of physical quantities of interest after 600 years (dashed line) and after 3900 years (dotted line). From top left in clockwise direction: density, temperature, v$_{\mathrm{rms}}$ and magnetic field intensity. The colors represent both uniform simulations, BLU (red) and BHU (blue).}
    \label{fig:evolution_histo_uniform}
\end{figure}

\begin{figure}
	\includegraphics[width=\columnwidth]{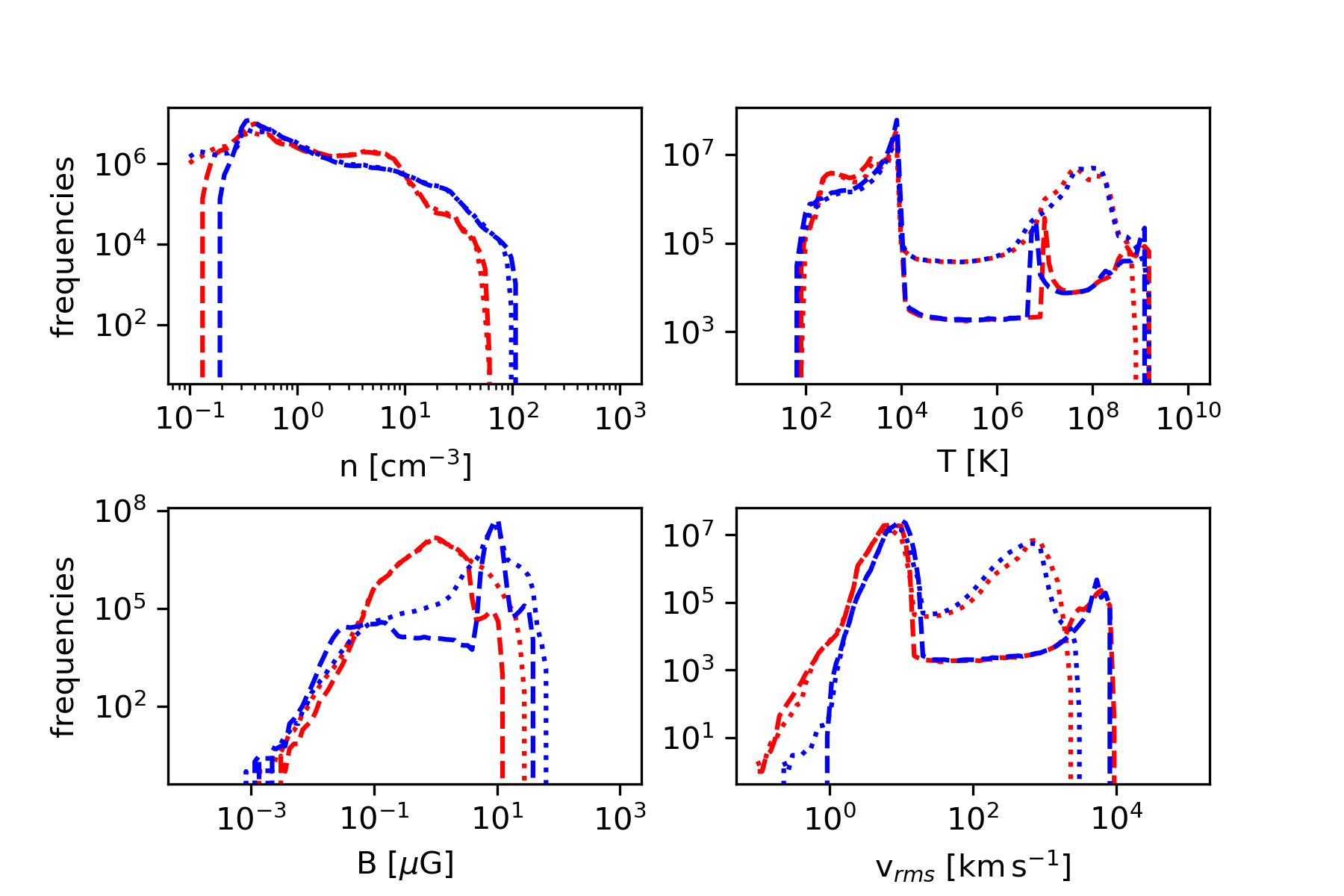}
    \caption{Histograms for the initial distributions of physical quantities of interest after 600 years (dashed line) and after 3900 years (dotted line). From top left in clockwise direction: density, temperature, v$_{\mathrm{rms}}$, and magnetic field intensity. The colors represent both turbulent simulations, BLT (red) and BHT (blue). }
    \label{fig:evolution_histo_turbulent}
\end{figure}

In this section we describe the evolution of four physical quantities that are of general interest in the physics of the ISM: density, temperature, velocity and magnetic field. Here we calculate histograms with the direct data from the simulations. We compare our four runs, the uniform runs BLU (red) and BHU (blue) shown in fig.\,\ref{fig:evolution_histo_uniform} and the turbulent ones BLT (red) and BHT (blue) displayed in fig.\,\ref{fig:evolution_histo_turbulent}. Both figures show two measurements taken at different time steps, at $t\sim600$\,yrs (dashed line) and at $t\sim3900$\,yrs (dotted line). We also calculated the angle between the velocity and magnetic field vectors for all the pixels that correspond to the SN shell. 

\begin{figure*}
	\includegraphics[scale=0.44]{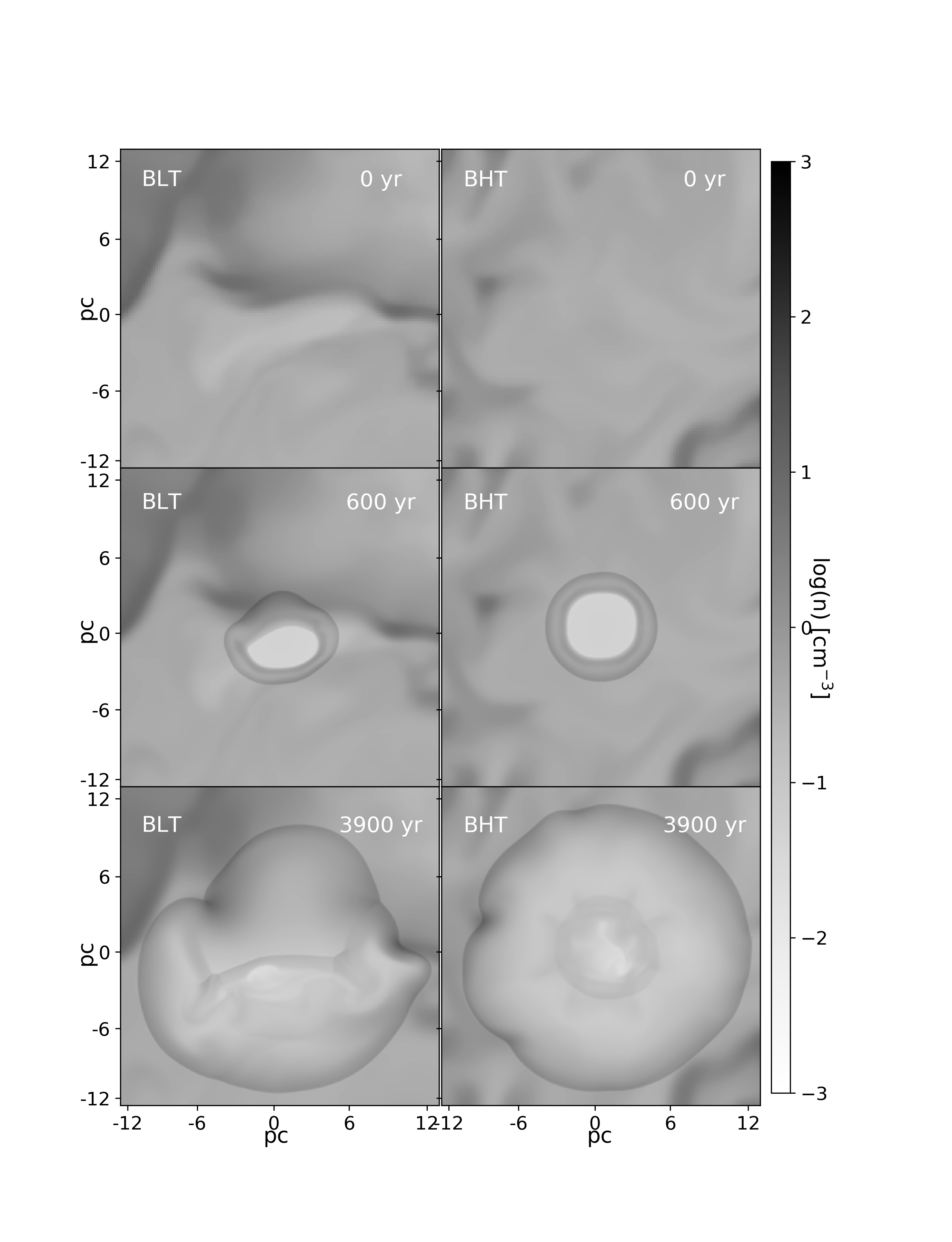}
    \caption{Density slices for the Turbulent simulations (top to bottom) at time 0, after 600 years and after 3900 years.}
    \label{fig:DensityMaps}
\end{figure*}

It is useful to look at the density slices of the SNR for the three selected time steps to get an idea of the physical distribution of the matter associated to the system of turbulent simulations, this can be seen in fig.\,\ref{fig:DensityMaps}. 

First, for the fiducial runs the density, temperature and magnetic field are uniform while the gas is initially at rest. This means that the widening of the distributions are entirely due to the effect of the expanding SNR. The only difference between these two runs is the magnitude of the initial magnetic field and this reflects on the magnetic field distributions (see bottom left of fig.\,\ref{fig:evolution_histo_uniform}). This implies that varying the uniform magnetic field does not alter the rest of the dynamical properties studied in this work, i.e. the density, temperature and velocity distributions are indistinguishable between them at both times. Additionally, the time necessary for the reverse shock to converge to the object's center is about 1100 years for both uniform runs. 

The SNR expanding in a uniform medium causes the density to increase by a factor of 4. The temperature histograms can be interpreted as the SNR expanding and cooling, leaving a heated portion of the gas behind, the velocity distributions starts as a delta that widens and slows down at later time steps. The magnetic field in both runs gets compressed increasing its maximum value by less than an order of magnitude. 

For the turbulent runs we looked at the scale of the system where the complete density distribution is not strongly altered by the evolution of the SNR. The maximum density of the gas does not appear to increase as the SNR advances through the medium, it either remains about the same as its initial maximum or slightly decreases at the latest time steps (see top left of fig.\,\ref{fig:evolution_histo_turbulent}). On the other end of the spectrum, the small density region ($n<10^{-1}$\,cm$^{-3}$) gets populated for both the BHT and the BLT cases. This gas is redistributed from the WNM peaks ($n \sim 2$\,cm$^{-3}$) in both models. 

The temperature distribution of the pre SNR ISM is not affected until the shock front enters in contact with it (top right in fig.\,\ref{fig:evolution_histo_turbulent}). The SNR introduces a delta distribution at very high temperature in a few central pixels, as it expands the distribution cools down, leaving an important population at high temperatures ($T > 10^8$\,K). The distribution for both BLT and BHT are very similar, the main difference between them being that BHT has a maximum slightly hotter than BLT.

The case of the velocity distribution is quite similar to the ones for temperature. The SNR is initially represented as a delta distribution at very high velocities compared to the velocities of the ISM, i.e initially there are two separated distributions (see bottom right of fig.\,\ref{fig:evolution_histo_turbulent}). As the SNR expands the velocity range between the two original distributions gets populated ($10^2$\,km s$^{-1} < v < 10^4$\,km s$^{-1}$). As the evolution of the system continues, the maximum velocity diminishes but the intermediate velocities reach higher population levels. 

The distributions that show the strongest changes are the ones for the magnetic field intensities (bottom left of fig.\,\ref{fig:evolution_histo_turbulent}). They are originally constrained to a small region of the spectrum and are widened along several orders of magnitude as the SNR expands. After 3900 years of SNR evolution the maxima of the magnetic field intensity distributions are enhanced by a factor of six for BLT and by about a factor of four for BHT with respect to their initial conditions. In contrast with the three physical variables aforementioned, the SNR does not introduce a magnetic field component, which means that the distributions shown always correspond to the magnetic field of the ISM. 

The maximum values for the density, magnetic field and velocity, after 3900 years for all the simulations are presented in table \ref{tab:final_values}.

\begin{table}
\caption{Table containing the maximum value of density, temperature, magnetic field and velocity for all the simulations after 3900 years.}
\centering
	\begin{tabular}{|c c c c|} 
 \hline
  Simulation & n [cm$^{-3}$] & B$_{\mathrm{rms}}$ [$\mu$G] & v$_{\mathrm{rms}}$ [km$\,$s$^{-1}$]\\ 
 \hline
 BLT & 58 &  26  & 2194\\ 
 \hline
 BHT & 92 & 62 & 2789\\
 \hline
 BLU & 7.45 & 4 & 604\\
 \hline
 BHU & 7.44 & 31 & 603\\
\end{tabular}
\label{tab:final_values}
\end{table}

\begin{figure}
	\includegraphics[width=\columnwidth]{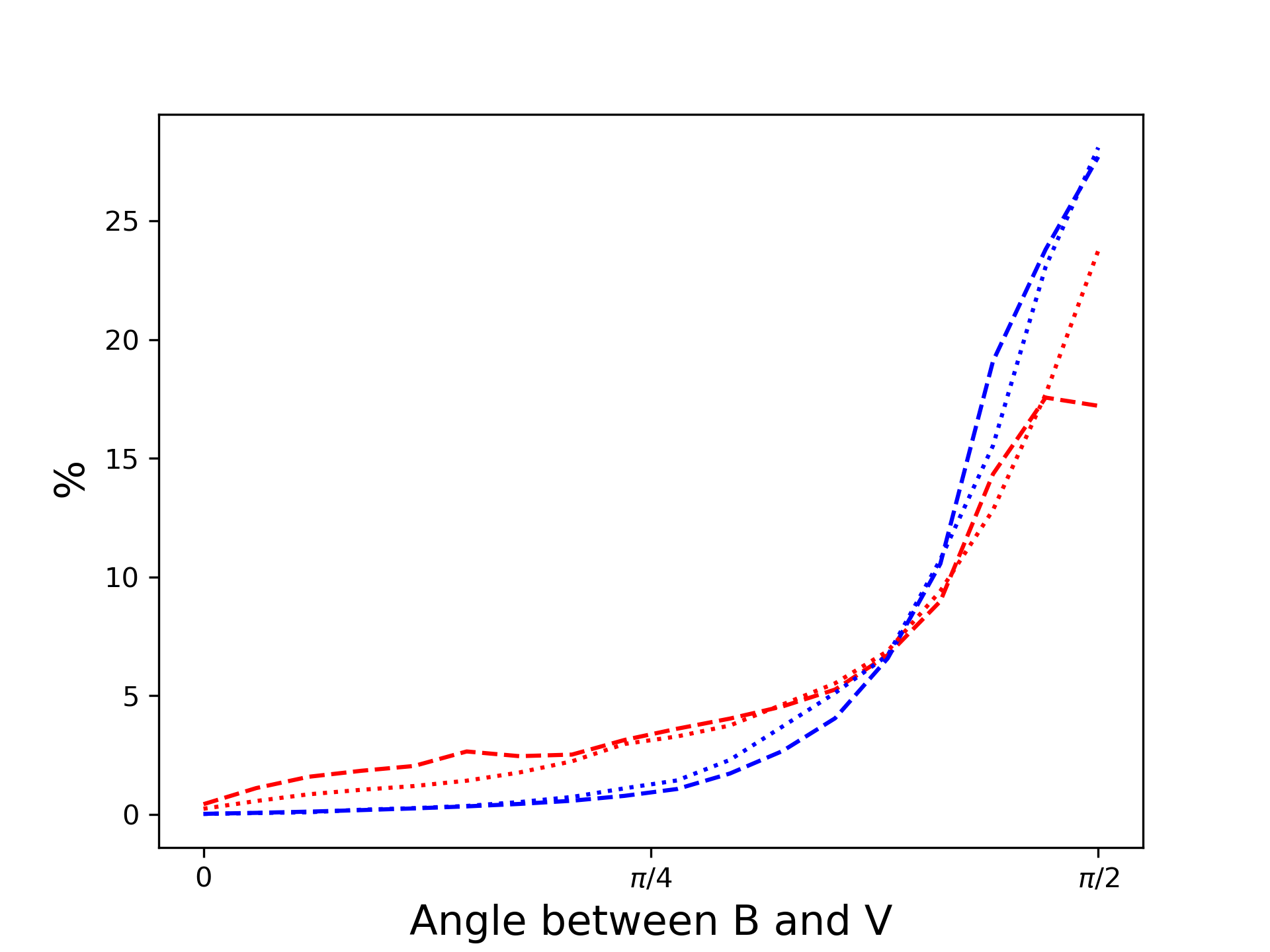}
    \caption{Histograms for the distributions of angles between $\vec{v}$ and $\vec{B}$ after 600 years (dashed line) and after 3900 years (dotted line). The colors represent both turbulent simulations, BLT (red) and BHT (blue).}
    \label{fig:Histogram_Angles_Turbulent}
\end{figure}

Finally, the angles between velocity and magnetic field at the SN can be seen in the histograms of figure \ref{fig:Histogram_Angles_Turbulent}. Those histograms were calculated for both turbulent runs and at the same time steps as the ones described before and share the same color and style schemes as the plots presented above. To ensure that we are extracting the pixels corresponding to the SNR shell we check that the gas that has not been in direct contact with the SNR remains static. This is a reasonable assumption considering the disparity in time scales for the evolution of the ISM ($\tau_{ISM} \sim 10^6$\,yrs) and for the SNR ($\tau_{SNR} \sim 100$\,yrs). In this scenario, we can calculate the integer ratio between the density fields at a certain time step and the initial condition. The result of this operation is a data cube with three possible values, a zero where the gas has been depleted, a one where the initial gas remains intact, and a number larger than one if the pixel corresponds to the shell of the SNR. This region is then used as a mask that selects the shell region in every data cube.  With the masked velocity and magnetic field cubes we can easily calculate the angle between two vectors by using eq.\,\ref{eq:cosine}. 

\begin{equation}
\theta = cos^{-1} \left( \dfrac{\vec{v} \cdot \vec{B}}{|\vec{v}| |\vec{B}|} \right)
\label{eq:cosine}
\end{equation}

The angle between magnetic field and velocity slightly differs between BLT and BHT. Both simulations show a clear preference for the vectors to be perpendicular and they mostly retain their distributions at different time steps. 

\subsection{Results from Synthetic Observations}

\begin{figure*}
	\includegraphics[scale=0.34]{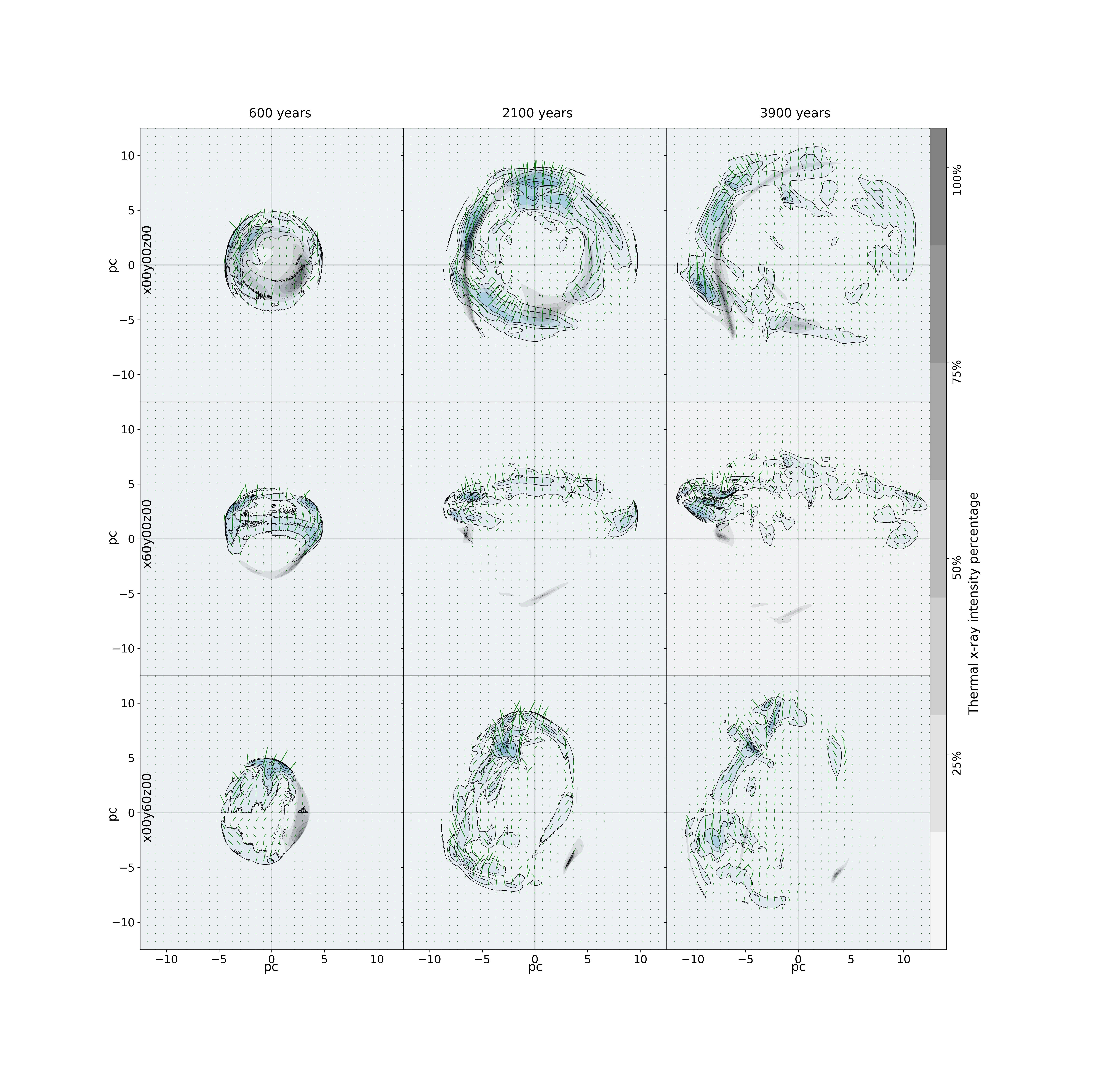}
    \caption{Time evolution of the simulated BLT SNR increasing in time from left to right. The black and grey contours represent polarization fraction intensity, the green lines are magnetic field vectors coming from polarization and the blue contours the x-ray emission strength fraction (per pixel, x-ray emission over maximum x-ray intensity of the image).}
   \label{fig:maps_BLT}
\end{figure*}

\begin{figure*}
	\includegraphics[scale=0.34]{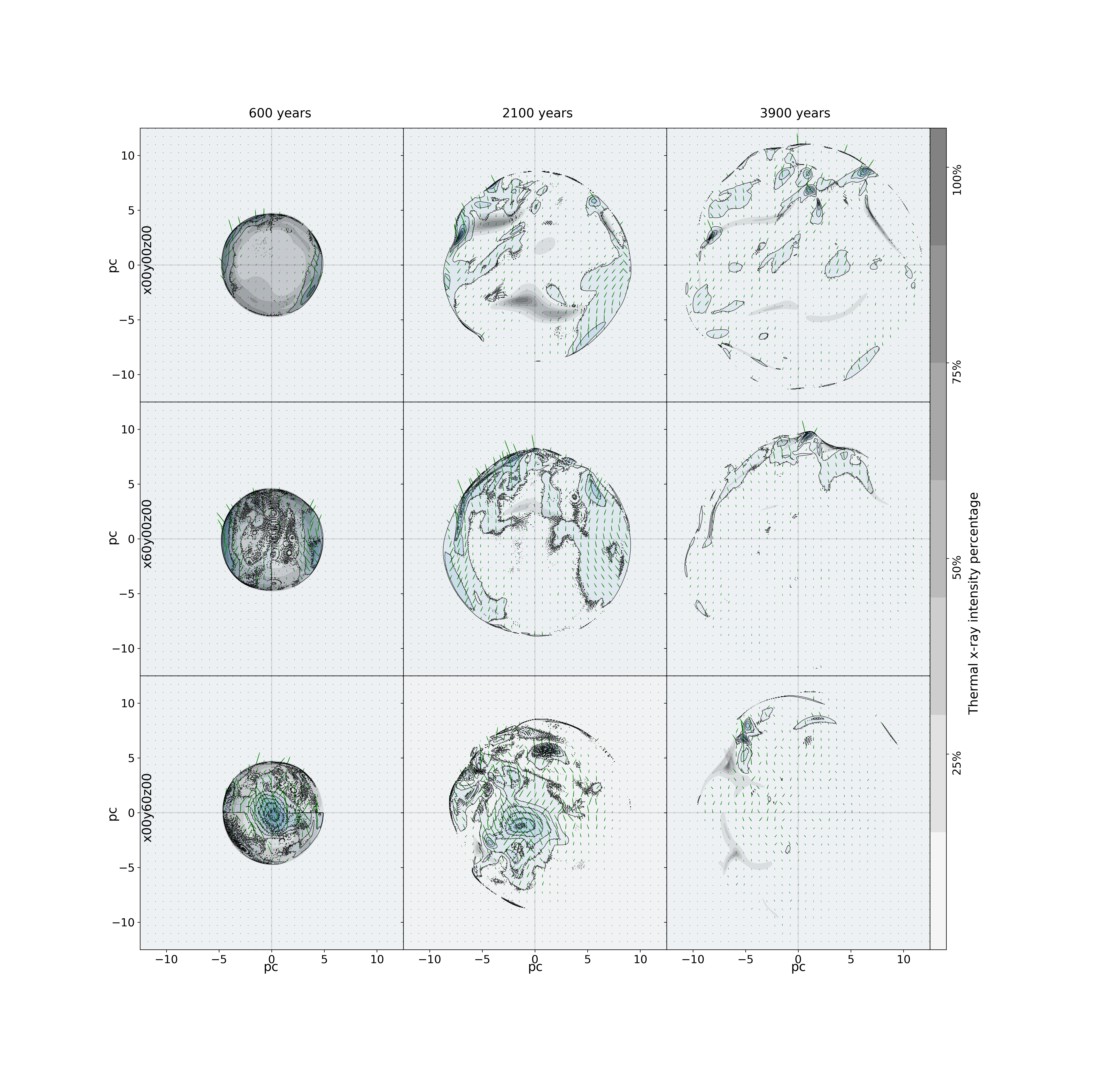}
    \caption{Time evolution of the simulated BHT SNR increasing in time from left to right. The black and grey contours represent polarization fraction intensity, the green lines are magnetic field vectors coming from polarization and the blue contours the x-ray emission strength fraction (per pixel, x-ray emission over maximum x-ray intensity of the image).}
    \label{fig:maps_BHT}
\end{figure*}

We studied the thermal X-ray emission (energy > 2\,keV) and the polarization of radio synchrotron emission (at 1.4\,GHz). Here we present the results of synthetically observing our SNR from three different points of view: non-rotated (x0y0z0), rotating the x axis by 60\textdegree (x60y0z0) and rotating the y axis by 60\textdegree (x0y60z0). The rotations are done in a frame of reference where the principal direction of the initial magnetic field is along the z axis. That axis is also the line of sight along which we integrated the data to generate our maps. These maps are shown in fig.\,\ref{fig:maps_BLT} (BLT) and fig.\,\ref{fig:maps_BHT} (BHT).

\begin{figure*}
	\includegraphics[width=\textwidth]{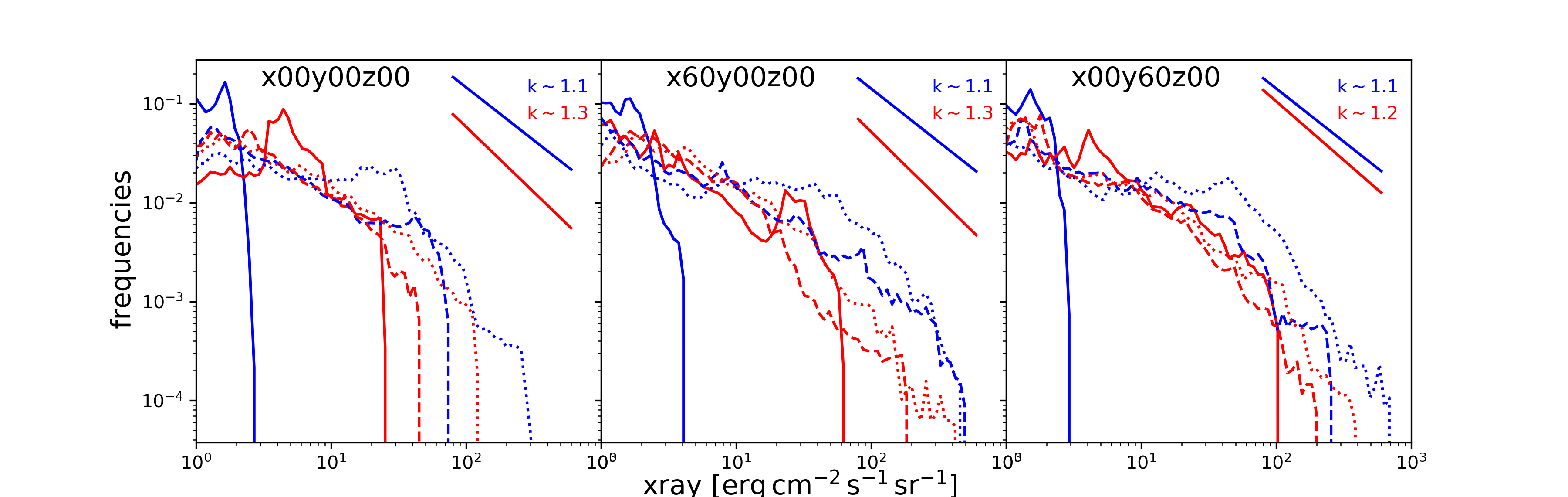}
    \caption{Thermal x-ray intensity after 600 years (dashed line) and after 3900 years (dotted line) for the two turbulent simulations BLT (red) and BHT (blue). The three panels represent different points of view of the supernova, from left to right: without rotation, rotated 60$^{\circ}$ along the x axis and rotated 60$^{\circ}$ along the y axis.}
    \label{fig:xray_histo}
\end{figure*}

In fig.\,\ref{fig:xray_histo} we show histograms of the thermal X-ray emission at two time steps: $t\sim600$\,yrs (dashed line) and $t\sim3900$\,yrs (dotted line). The simulation BHT reaches the highest energies in the three points of view, and also has higher counts in most of the energy bins. If we approximate the distributions at $t \sim 3900$\,yrs as power laws\footnote{The power law exponents were calculated using the tool described in \cite{Alstott14}} ($frequencies \propto energy^k$) there is a small difference between the slopes of the different runs; BHT has a similar slope for the projected planes studied while BLT has faintly different slopes that are steeper, compared to BHT, in the all three points of view.

\begin{figure}
	\includegraphics[width=\columnwidth]{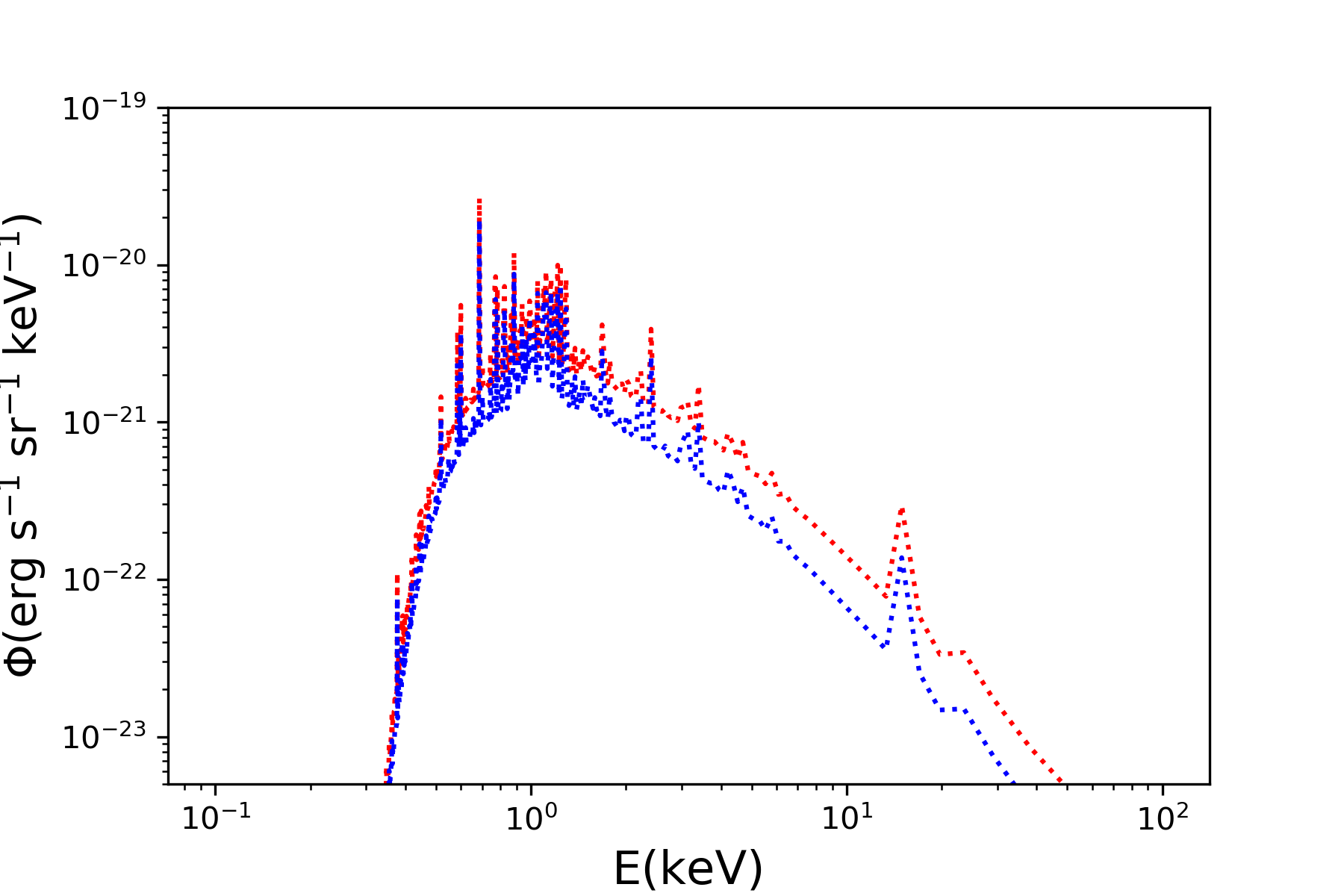}
    \caption{Thermal x-ray spectrum for BLT (red) and BHT (blue) after 3900 years}
    \label{fig:xray_spectra}
\end{figure}

In fig.\,\ref{fig:xray_spectra} we show the spectra for the thermal X-ray after 3900 years. In said figure it can be seen that both simulations render quite similar spectra, but BLT is always more intense than BHT.

The BLT simulation shown in fig. \ref{fig:maps_BLT} at 3900 yrs looks like a shell remnant that may be spherical, bilateral or highly deformed depending on the point of observation, in the other hand the BHT ones at the same time appear to be a shell remnant that is broken to different degrees. 

After 3900 years the regions where radio emission is more intense do not coincide with the most intense x-ray emission regions for the three points of view of both SNRs. 

\begin{figure*}
	\includegraphics[width=\textwidth]{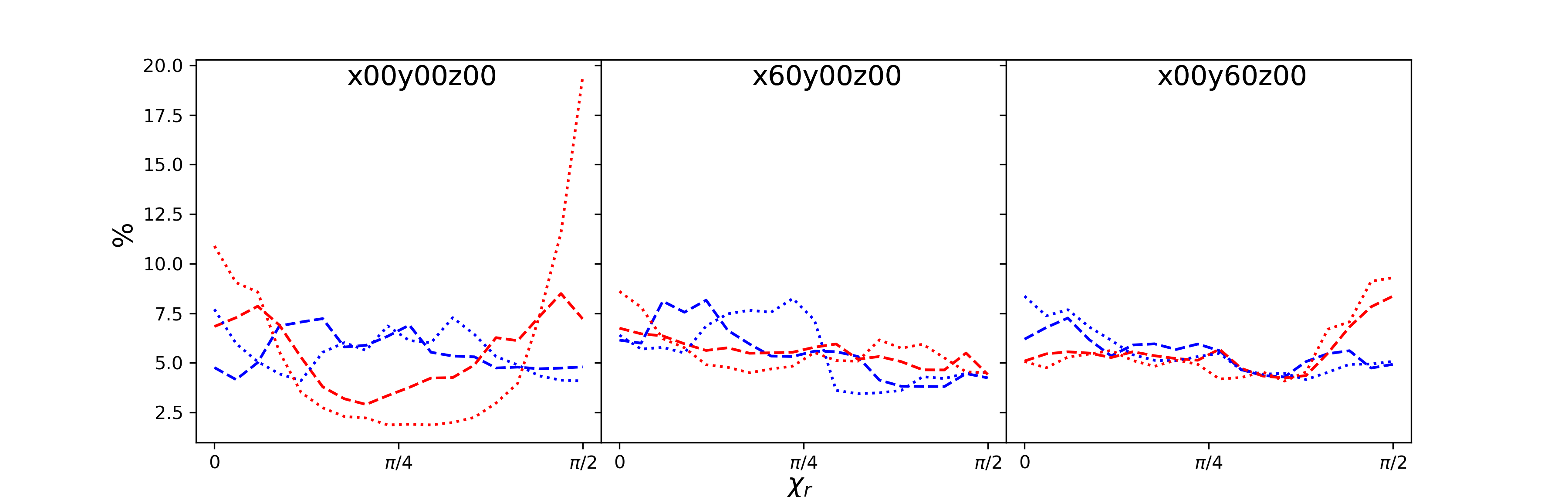}
    \caption{Polar referenced angle distribution of the magnetic field after 600 years (dashed line) and after 3900 years (dotted line) for the two turbulent simulations BLT (red) and BHT (blue). The three panels represent different points of view of the supernova, from left to right: without rotation, rotated 60$^{\circ}$ along the x axis and rotated 60$^{\circ}$ along the y axis.}
    \label{fig:pol_ang_obs}
\end{figure*}

The most intense radio emission zones coincide with the strongest polarization vectors. These vectors do not have an apparent preferred direction, specially considering that the point of view does alter the direction of the vector with respect to the shell. This is quantified in fig. \ref{fig:pol_ang_obs}, where we plot histograms of the polar referenced angle\footnote{A polar referenced angle is the angle that exist between a given vector and a vector pointing to the center of a circle.} distribution of the magnetic field in the SNR shells. This figure shows that for the same object the magnetic field
may be seemingly parallel, perpendicular or without any preference depending just on the point of observation. The magnetic field magnitude does affect how this angle is perceived as BLT has stronger changes between points of observation, in contrast to BHT which has few and small changes in their histograms. 

\section{Discussion}
\label{sec:Discussion}

We have studied the expansion of a SNR in a turbulent, thermally bi-stable and atomic ISM. We carried out this study by using three dimensional MHD simulations in 25\,pc$^3$ boxes. The initial magnetic fields of the simulations have an important uniform component. We tested two scenarios with the strongest difference between them being the average magnitude of the magnetic field, one staying close to 4\,$\mu$G (BLT) while the other one at 15\,$\mu$G (BHT). We let the remnants evolve and expand for a few thousand years, we created maps through synthetic observations and analyzed the results. 

The SNR expands faster in the regions where the density is lower. In our simulations the denser regions consist of filament-like clumps. There have been previous publications that tackle the phenomenology of a hydrodynamic or MHD shock interacting with ISM clouds, for example in \cite{PittardParkin16} the authors studied the interaction of a three-dimensional system of a shock front and a spherical clump with soft edges that is in pressure equilibrium with a uniform background, the presence of soft edges in the clouds hinters the development of Kelvin-Helmholtz (KH) like instabilities. Similarly in \cite{PittardGoldsmith16} the clumps have a cylindrical (filament like) shape while \cite{GoldsmithPittard16} uses the MHD equations to study the interaction of shocks and filaments. The previously mentioned works report that the rate of destruction and redistribution of the cloud's mass is influenced by the orientation of the filament's main axis and the shock and the magnetic field, this is specially relevant for clumps that have a size similar to the size of the shock. In our case, if the SNR encounters a cloud after a few hundred years, it does not generally break the filaments but it's mass is in fact redistributed. The remnant grows surrounding the obstacle, similar to the experimental setups aimed at generating turbulent vortices. The turbulent amplification of the magnetic field is strongly associated to instabilities like KH or the Richtmyer-Meshkov instabilities. Those instabilities manifest themselves as sub-structures where the amplification is stronger. In our case the resolutions of our simulations and the HLLD scheme employed in our code (which is very diffusive) do not allow to study in detail the amplification processes, nevertheless the magnetic field is in fact amplified, mostly due to rearranging or piling up the ISM's magnetic field and it does reach values that are quite higher than the initial maxima and are also higher than the standard value of 6\,$\mu$G of the ISM \citep{Beck08}. Higher resolution simulations of SNR expansion have been tested in the two dimensional case by several groups, for example \cite{Inoue09} and \cite{Guo12} achieved amplifications of more than an order of magnitude with their turbulent simulations. These kind of works tend to report the highest magnetic field amplification localized in the small scale substructures associated to MHD instabilities. After 3900 years of evolution our maximum magnetic fields are not exorbitantly high, as they only reach 26 $\mu$G (BLT) and 62 $\mu$G (BHT). However our simulations do show an important increase in the magnetic energy as the SNR evolves. Our BLT simulation increases its magnetic energy by 34$\%$, while BHT only increases by 4$\%$. Evolving in a non-uniform medium favors the amplification of magnetic energy and from our simulations we see that starting with a less intense magnetic field favors the occurrence of dynamo effects which increase the magnetic energy, this may point to the role of magnetic tension in reducing the SNRs shock front impact on the ISM.
As the SNR expands it starts cooling via radiation. The most important kind of radiation is either free-free or free-bound emission, both of which are proportional to $n$. This means that for denser clumps encountered by the shock front the stronger the emission produced. Hence, the SNR encountering a filament should be easily detected in observations, both real and synthetic. This result is sufficient to explain the cooling of the hottest points of the temperature histograms as the SNR advances as well as the increase in the x-ray emission energies (fig.\,\ref{fig:evolution_histo_turbulent},fig.\, \ref{fig:xray_histo}). Both simulations, from the three points of view, show fairly similar slopes in their thermal x-ray emission spectrum. However BHT, which reaches higher densities of up to $92$\,cm$^{-3}$, is above BLT, which stays with a maximum of $56$\,cm$^{-3}$, in most of the thermal x-ray energy ranges.  

The magnetic field of the ISM is important for the synchrotron radiation of SNRs both in its amplified or in its pre-contact state. In fact, works like the ones by \cite{Reynoso18} argue that the ISM magnetic field may be detectable even after being in contact with a SNR shock front. Simulations have also been used to study the orientation or reorientation of the magnetic field of the ISM after the passage of a SNR, for example \cite{Bao18} simulated the expansion of a SNR in a media with a magnetic field with an ordered component finding that different lines of sight produce different polarization maps. These authors found stronger polarization intensities when observed along the ordered component of the magnetic field and lower values when deviated from that angle. Additionally, they report that the magnetic field seems to be mostly radial when the line of sight coincides with the ordered magnetic field but less clear when observed from other points. Our histograms for the three-dimensional values of the SNR show that if the shell has velocities mainly in the radial direction, the magnetic field is predominantly perpendicular to the SNR radii and that this trend increases over time (fig.\,\ref{fig:Histogram_Angles_Turbulent}). As mentioned before this can be understood by considering that the magnetic field piles up perpendicularly to the shock front. Observing magnetic field in the radial direction is considered to be a consequence of Rayleigh-Taylor instabilities in the shock front in young SNRs (starting the Sedov-Taylor phase of evolution). Since the SNRs shown in this work are middle-age objects, i.e., they are in a really advance stage of the Sedov evolution phase, we anticipate a low population of radial magnetic fields in our histograms. 

 Evaluating our synthetic maps; the angles between radial vectors and the magnetic vectors in them results in polar referenced angle histograms (fig.\,\ref{fig:pol_ang_obs}). In those histograms a peak near $\pi /2$ means that magnetic field is tangential to the shock normal, like in the three dimensional case. When the maximum of the histogram is near $\pi /4$ it represents an intermediate state in which the magnetic field is neither parallel nor perpendicular to the shock normal. A third scenario encountered is a flat histogram, where every angle has similar populations which means that we can have the same amount of vectors parallel and perpendicular. 

Those are the scenarios we encountered in our results, they appear in different order as we changed the point of view. None of our tested scenarios shows a radial magnetic field. In fact the region of angle $0$ is underpopulated in most histograms. For the three point of view of the BHT simulations, we find that after 600\,yr or 3900\,yr show a relatively flat histogram with a slight preference for smaller angles. When the line of sight is in the direction of the uniform magnetic field, BLT shows a strong preference for its magnetic field to be tangential to the shock normal specially after 3900\,yr. Specifically this histogram also shows an important population of radial magnetic field. However those vectors are not associated to the strongest emitting regions (fig.\,\ref{fig:maps_BLT}).

\section{Conclusions}
\label{sec:Conclusions}

In the present work we performed three-dimensional simulations of SNR expanding on a magnetized turbulent neutral atomic ISM. We analyzed three-dimensional results and also created two-dimensional maps from synthetic observations on radio and x-ray emission. The main conclusions from our results are: 

\begin{itemize}

  \item The use of an atomic ISM segregated via TI as the background of three-dimensional simulations of SNR expansions offers a new insight into the kind of turbulent media that surrounds SNR; 
  
  \item The strongest changes to the ISM due to the interaction with a SNR are in temperature, velocity and magnetic field. Density is not strongly affected; 

  \item The ISM's magnetic field together with the expansion of the SNR achieves an increase of the magnetic energy of 34\,$\%$;

  \item The magnetic field is clearly perpendicular to the SNR shell in three-dimensions. However, the apparent orientation of the magnetic field in two-dimensional maps changes drastically, even showing the opposite behaviour, making it highly difficult to determine the true nature of the ISM's magnetic field from these maps. 
  
  \item The results in this paper are an improvement on the current knowledge of how magnetic field is amplified and redistributed in the atomic ISM. Similar simulations with higher resolutions will undoubtedly shed new light into this important topic.

\end{itemize}

\section*{Acknowledgements}

M.V. is doctoral fellow of CONICET, Argentina. D.G. and E.G. are members of the Carrera del Investigador Cient\'{i}fico of CONICET. This work was partially supported by Argentina grants UBACYT 200201501000098BA, ANPCYT(PICT 0571/11) and ANPYCT(PICT 1707/15). PFV acknowledges financial support from PAPIIT IG100218 grant (UNAM, Mexico). 





\bibliographystyle{mnras}
\bibliography{ref} 



\bsp	
\label{lastpage}
\end{document}